# Monthly sunspot number time series analysis and its modeling through autoregressive artificial neural network

[1]Goutami Chattopadhyay, [2†]Surajit Chattopadhyay
[1]Department of Mathematics, Bengal Engineering and Science University, Shibpur, Howrah 711103
[2]Pailan College of Management and Technology, Kolkata 700 104
[2]Author for correspondence
Email: surajit_2008@yahoo.co.in, surajcha@iucaa.ernet.in
[2†] Visiting Associate of Inter-University Centre for Astronomy and Astrophysics (IUCAA), Pune, India

**Abstract**

This study reports a statistical analysis of monthly sunspot number time series and observes non homogeneity and asymmetry within it. Using Mann-Kendall test a linear trend is revealed. After identifying stationarity within the time series we generate autoregressive AR(p) and autoregressive moving average (ARMA(p,q)). Based on minimization of AIC we find 3 and 1 as the best values of *p* and *q* respectively. In the next phase, autoregressive neural network (AR-NN(3)) is generated by training a generalized feedforward neural network (GFNN). Assessing the model performances by means of Willmott's index of second order and coefficient of determination, the performance of AR-NN(3) is identified to be better than AR(3) and ARMA(3,1).

## 1.    Introduction:

Sunspots are generated by strong magnetic fields that are created in the interior of the Sun. The sunspots are basically visible surfaces within the Sun that are pock-marked with black flecks. Of all solar features, the sunspots are the most easily observed and their interemmittent appearance has numerous impacts on Earth. The sunspots may appear as single, isolated umbra (the dark central region) surrounded by a symmetric penumbra (less dark pattern surrounding umbra) or they may appear in groups. The process of creation of sunspots due to the solar magnetic fields is thoroughly discussed in Noyes (1982). It has been pointed out long back that the records of sunspot appearances indicate a strong tendency of having an 11-year cycle. The origin of the sunspot cycle has been an area of research since long. Central to the occurrence of the 11-year cycle is the oscillating magnetic dynamo within the Sun. Sunspot cycles are observed to vary both in size and length; therefore, it is difficult to describe the shape of sunspot cycles with a universal function (Lasheng et al, 2005). Plethora of literatures is available where the authors have attempted to describe the cycles as a periodic phenomenon (e.g. Denkmayr and Cugnon, 1997; Hanslmeier et al, 1999; Wang et al., 2002; Zhan et al., 2003; Zhou et al., 2002; Xu et al, 2008). Understanding the solar activity cycle remains a key unsolved problem in solar physics (along with, e.g., heating of the solar corona and solar flares). It is not only an outstanding theoretical problem, but also an important practical issue, since solar activity and the radiation output related to it influences the biosphere, space weather and technology on the Earth (Brajša et al, 2009; Hanslmeier, 2007). In the past, it has been postulated that the irregular dynamics of the solar cycle may embed a low order chaotic process (Spiegel, 1994) which, if true, implies that the future behaviour of solar activity should be predictable (Orfila et al, 2002). A new concept of the solar cycle as a low-dimensional chaotic system was introduced by Ruzmaikin (1981), and since early 1990's, many authors have considered solar activity as an example of low-dimensional deterministic chaos, described by strange attractor (e.g., Usoskin and Mursula, 2003 and references therein). A dynamic system approach to the prediction of solar cycle was adopted by Orfila et al (2002).

Sunspot number (SN) is the most commonly predicted solar activity index. The rate of solar flares and amount of energy they release are well correlated with the sunspot number, as is the rate of coronal mass ejections. Cosmic rays, whose flux is anticorrelated with the solar cycle, are a significant source of radiation hazard in space. Geomagnetic activity has one component that is proportional to SN and another, which can be a source of significant space weather that resembles the sunspot number but shifted forward several years (Pesnell, 2008). Categorized prediction of solar cycle has been thoroughly reviewed in Pesnell (2008). Impact of solar cycle on climate variability has been discussed in Shindell et al (1999), which shows from a global climate model including an interactive parameterization of stratospheric chemistry how upper stratospheric ozone changes may amplify observed, 11-year solar cycle irradiance changes to affect climate. The effect of solar cycle on climate has also been reviewed by Varotsos and Cracknell, (2004), Varotsos (2002) and Varotsos (1989). Various statistical methods of predicting solar cycle have been reviewed in Hathaway et al (1999). Petrovay (2011) presented an exhaustive review work on solar cycle prediction. In the review, Petrovay (2011) categorized the prediction methods form three main groups, namely, precursor methods, extrapolation methods and model based predictions. Attractor analysis and phase space reconstruction methods have been reviewed in Petrovay (2011) and this includes artificial neural network (ANN) as a non-parametric fitting to the sunspot number time series. Detailed theoretical accounts of ANN are presented in Rojas (1996). Inspired by biological systems, particularly by research into the human brain, ANNs are able to learn from and generalize from experience (Zhang et al, 1998). An extensive discussion on the suitability of ANN in complex forecasting problems is available in a review by Zhang et al (1998). Silverman and Dracup (2000) summarized the advantages of ANN over conventional statistical methods as:

- a priori knowledge of the underlying process is not required;
- existing complex relationships among the various aspects of the process under investigation need not be recognized;
- constraints and a priori solution structures are neither assumed nor enforced.

Since Yule (1927) first proposed a technique called an autoregressive method, a great variety of techniques have been proposed to predict the magnitude of sunspot activities (e.g. Zhan et al, 2003; Obridko and Shelting, 2008). ANN methods, till date, have been used by some authors to predict the sunspot activities (e.g. Verdes et al, 2000; Podladchikova and Van der Linden, 2011; Obridko and Shelting, 2008; Calvo et al, 1995). Various numerical prediction techniques have been used for the sunspot number time series, e.g. curve fitting, artificial intelligence, neural networks and so on (Xu et. al., 2008).

The present paper is aimed at predicting the monthly sunspot number (SN) in a univariate manner. The paper is organized as follows: First we have examined the time series for presence of any homogeneity and linear trend by means of Pettitt's test and Mann-Kendall trend analysis respectively. Then the monthly SN time series has been examined for symmetry by means of Kolmogorov-Smirnov test. The tests stated above have been aimed at putting some light into the intrinsic behaviour of the monthly SN time series. Subsequently, the time series have been investigated for the presence of any non-stationarity by means of fitting autoregressive model and a comparative study by means of minimization Akaike Information Criterion (AIC) has been made among usual autoregressive AR(p) and autoregressive moving average (ARMA(p,q)) methods. The order of autoregression has been identified in this manner and based on this autoregressive neural network (AR-NN) has been trained to investigate its suitability in predicting monthly sunspot numbers.

## 2. Methodology:

Methodology of this work consists of
- Pettitt's test for homogeneity
- Mann-Kendall trend analysis
- Kolmogorov-Smirnov test for normal distribution
- Autoregressive modeling
- Development of ANN
- Skill assessment of the prediction model

### 2.1 Pettitt's test for homogeneity

Pettitt's test (Pettitt 1979, Demarée 1990, Nordli 1996, Karabork et al, 2007) is used to investigate the homogeneity of a long time series. In geophysical time series analysis, this technique has been adopted by several authors. The test statistic $X_k$ is defined by (Rutishauser et al, 2009)

$$X_k = 2R_k - k(n+1)$$

$$R_k = \sum_{i=1}^{k} r_i$$

where $r_i$ is the rank of the ith element in the complete series of n elements. If shifts are absent in the series, i.e. under the null hypothesis of randomness, the expectation value of $X_k$ is 0.

## 2.2 Mann-Kendall test for trend

Mann-Kendall (MK) test is a very popular tool for identifying the existence of increasing or decreasing trend within a time series. The MK test is the rank-based nonparametric test for assessing the significance of a trend, and has been widely used in climatological trend detection studies. Examples include the studies of Hanssen-Bauer and Forland (1998), Shrestha *et al.* (1999), Domonkos *et al.* (2003) and Chattopadhyay et al (2011). The null hypothesis $H_0$ is that a sample of data $\{Y_t : t = 1, 2, \ldots, n\}$ is independent and identically distributed. The alternative hypothesis $H_1$ is that a monotonic trend exists in $\{Y_t\}$. Each pair of observed values $(y_i, y_j)$, where $i > j$ is inspected to find out $Y_i > Y_j$ (first type) or $Y_i < Y_j$ (second type). There is a correction for the case $Y_i = Y_j$. If the numbers of first type and second observations be *P* and *M* respectively, then the statistic *S* is defined as $S = P - M$. A standard normal variate $Z$ is now constructed following Yue *et al.* (2002). In a two-sided test for the trend, the null hypothesis of no trend is rejected if $|Z| > Z_{\alpha/2}$, where α is the significance level.

## 2.3 Kolmogorov-Smirnov test for normal distribution

The Kolmogorov-Smirnov (KS) statistic provides a means of testing whether a set of observations are from some completely specified continuous distribution, The KS test has two major advantages over the chis-quare test that is an alternative for this purpose :

    1. It can be used with small sample sizes, where the validity of the chisquare test would be questionable.

    2. Often it appears to be a more powerful test than the chi-square test for any sample size.

KS test for normal distribution has been discussed thoroughly in Lilliefor (1967). The KS test has been carried out for climatological studies by several authors. Examples include Gershunov et al (2001), Jamaludin and Jemain (2007) and Husak et al (2007).

The KS statistic, given by *D*, is based on the largest vertical difference between the empirical distribution function $F_n(x_{(1)})$ and the hypothesized distribution function $F(x_{(i)}, \hat{\theta})$. The computational formula is given by

$$D^- = \max_i \{Z_{(i)} - (i-1)/n\}$$
$$D^+ = \max_i \{i/n - Z_{(i)}\}$$
$$D = \max(D^+, D^-)$$

Here, $Z_{(i)}$ represents an ordered dataset with $F(x_{(i)}, \hat{\theta})$. In this statistical test, the null hypothesis is that the observed data are drawn from the chosen theoretical distribution. If the value of the KS-statistic is excessively large, then the null hypothesis is rejected. A rejection would imply that the distribution parameters are not doing an adequate job of modeling the empirical distribution of rainfall at a location. The acceptable KS value for rejection depends on the number of points in the empirical distribution being used to test the theoretical distribution, and the rejection level chosen for the study (Husak et al, 2007).

## 2.4 Autoregressive modelling

Autoregressive models have long been used in astronomical data analysis. Vio et al (2005) reviewed various potentialities and limitations of astronomical time series data analysis where the autoregressive model is discussed as a discrete approach. Buffa and Porceddu (1997) developed autoregressive neural network for temperature forecast and dome seeing minimization. In a more recent work, Brajša et al (2009) described autoregressive moving average (ARMA) for solar cycle predictions and reconstructions. The set of adjustable parameters $\phi_1, \phi_2, \ldots, \phi_p$ of an autoregressive process of order p, i.e. AR(p) process (Box *et al.*, 2007)

$$\tilde{z}_t = \phi_1 \tilde{z}_{t-1} + \phi_2 \tilde{z}_{t-2} + \ldots + \phi_p \tilde{z}_{t-p} + a_t$$

satisfies certain conditions for the process to be stationary. Here, $\tilde{z}_t = z_t - \mu$. The parameter $\phi_1$ of an AR(1) process must satisfy the condition $|\phi_1| < 1$ for the time series to be stationary. It can be shown that the autocorrelation function satisfies the equation

$$\rho_k = \phi_1 \rho_{k-1} + \phi_2 \rho_{k-2} + \ldots + \phi_p \rho_p$$

Substituting $k = 1, 2, \ldots, p$ in the above equation we get the system of Yule-Walker equations (Box *et al.*, 2007)

$\rho_1 = \phi_1 + \phi_2 \rho_1 + \ldots + \phi_p \rho_{p-1}$

$\rho_2 = \phi_1 \rho_1 + \phi_2 + \ldots + \phi_p \rho_{p-2}$

$\vdots$

$\rho_p = \phi_1 \rho_{p-1} + \phi_2 \rho_{p-2} + \ldots + \phi_p$

The Yule-Walker estimates of the autoregressive parameters $\phi_1, \phi_2, \ldots, \phi_p$ are obtained by replacing the theoretical autocorrelation $\rho_k$ by the estimated autocorrelation $r_k$. Thus, the matrix notation, the autoregression parameters can be written as:

$\Phi = R^{-1} r$

The $p$th-order autoregressive process may be written as

$\phi(B)\tilde{z}_t = e_t$

Where, $e_t$ follows the $q$th-order moving average process

$e_t = \theta(B) a_t$

Now, an ARMA($p, q$) process is presented as

$\phi(B)\tilde{z}_t = \theta(B) a_t$

where $\phi(B)$ and $\theta(B)$ are polynomials of degree $p$ and $q$ respectively, and $B$ is the backward shift operator. The ARMA process is stationary if the roots of $\phi(B) = 0$ lie outside the unit circle and it exhibits explosive non-stationary behaviour if they lie inside the unit circle.

## 2.5 Development of artificial neural network (ANN)

The method of choosing the predictors in the univariate problem under consideration would be based on the outcomes of the experiments based on the methodologies explained above. A generalized feedforward neural network (GFNN) would be adopted with two hidden layers. The number of nodes in both of the hidden layers is 4. In both of the hidden layers the sigmoid nonlinearity is used. It should be stated that before applying the ANN methodology we have removed the seasonal and trend components from the data. Trend and seasonality removal processes are referred as pre-whitening methods (Taskaya-Temizel and Casey, 2005). The training and test dataset ration is 3:1. The GFNNs are a generalization of the conventional multilayer perceptron such that connections can jump over one or more layers (Arulampalam and Bouzerdoum, 2003; Bouzerdoum and Mueller, 2003). Theoretical and mathematical details are available in Rojas (1996). The GFFN model is trained in batch mode up to 3000 epochs and run thrice. Minimization of the mean squared error is considered as the stopping criterion. The implementation details are described in the subsequent section.

## 3. Discussion:

In this work, the time series of monthly SN for the years 1992-2008, have been analyzed for various statistical aspects. First of all, homogeneity within the data set has been investigated through Pettitt test where, null hypothesis has been assumed that the data are homogeneous. The two tailed p-value computed using Monte Carlo simulation shows that the p-value is less than 0.0001. Since the computed p-value is found to be lower than the significance level $\alpha(=0.05)$ the null hypothesis is rejected and hence it may concluded that the data are not homogeneous. After identification of inhomogenety within the data set, we examine the data set for simetry and it is found that the time series of monthly SN is positively skewed. For further confirmation of asymmetry we carry out Kolmogorov-Smirnov test to examine a null hypothesis that assumed that the data are following a normal distribution. However, the computed p-value 0.005 (<0.05), it is concluded that the time series under consideration does not follow any normal distribution.

Subsequently, we apply Box-Cox transformation for smoothing of the time series. The $\lambda$ parameter for this transformation is optimizing to 0.345. The data series after Box-Cox transformation is presented in second panel of figure-01 which shows reduce variability within the time series plotted in the first panel of figure-01.

In the following step, we explore the transformed time series for existence of any linear trend within it for this purpose. We carry out M-K test (two tailed) based on the null hypothesis of no trend within the time series. Kendall $\tau$ is found to be -0.205 with $S = -4169$. The two tailed p-value being found to be less than 0.05, the null hypothesis rejected against the alternative hypothesis assuming the existence of linear trend. Hence it is concluded that the time series is characterized by a linear trend within it. Therefore, the statistical analysis carried out so far identifies the following:

- The monthly SN time series is inhomogeneous
- The time series is not symmetric and positively skewed
- There is significantly variability within the time series
- The time series is characterized by linear trend.

In the next step, it is necessary to identify a representative statistical model for said in homogeneous time series, as we are working in a univerient environment, an autoregressive approach may be adopted to generate a representative statistical model for this time series. The basic question at this juncture that requires to be answered is whether the time series is stationary. To solve this problem we fit a first order autoregressive (AR1) model to this time series. An AR equation is basically a linear non homogeneous recurrence relation. For the present problem fitted AR (1) model is given by $X_{t+1} = 0.931 X_t + 54.449$, whose, characteristics root lies within the unit circle. Hence, the sunspot number time series under consideration is stationary. Subsequently we carried out AR models for higher orders and based on the minimization of the AIC, the AR(3) is considered as better than other orders. To further improve the model we adopt ARMA models with autoregression order 3 and variable orders of moving averages. Minimization of AIC indicates ARMA(3,1) as better than others, that is, increasing the order of the moving average is not increasing the prediction performance of the model. The AICs are presented as a bar diagram in figure 1.

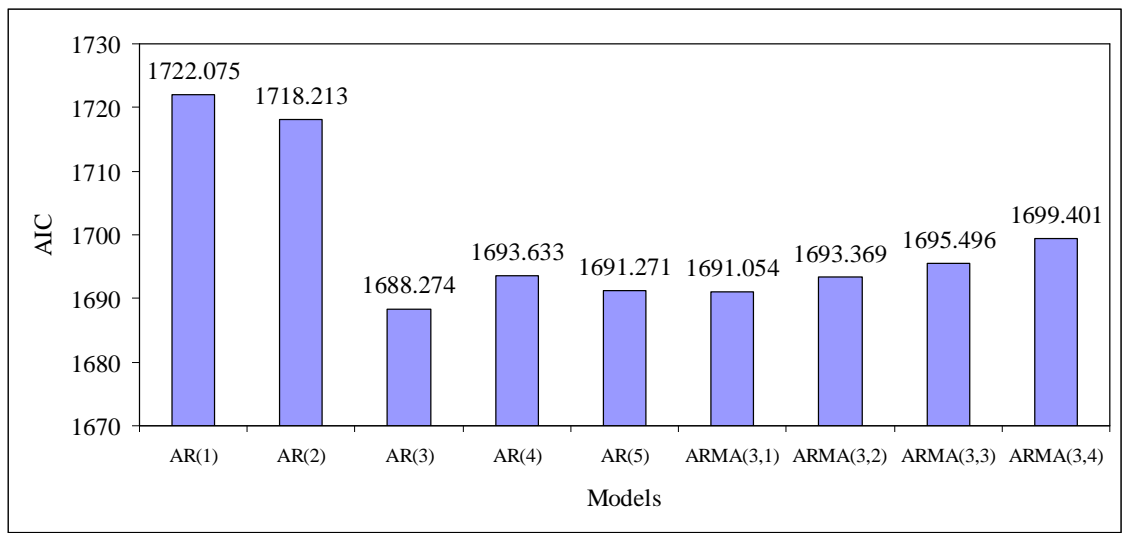

Figure 01. The values of AIC for different models

In the next step, our target is to examine whether the prediction performances of AR(3) and ARMA(3,1) may be enhanced by means of AR-NN(3), i.e. autoregressive neural network with three predictors. The predictors are $X_{t-3}, X_{t-2}, X_{t-1}$ for the predictand $X_t$, where the suffix indicates the time of observation. Hence, the previous three consecutive values of the same SN time series are being used to predict the current value of the same time series by means of neural network. That is why it is being dubbed as "autoregressive neural network" with three predictors and being abbreviated as AR-NN(3). The entire dataset being consisted of 204 monthly SN data, the input matrix is of order $(201 \times 3)$ and the output matrix is of order $(201 \times 1)$. The entire dataset is divided into training and test cases in the ratio 3:1. Hence, the number of test cases is 51. The model is tested over these 51 test cases.

To asses the model performance we consider Willmott's index advocated by Willmott (1982) as an index to measure the degree of agreement between actual and predicted values. This is given as:

$$d^2 = 1 - \left[\sum_i |P_i - O_i|^\alpha\right]\left[\sum_i \left(|P_i - \overline{O}| + |O_i - \overline{O}|\right)^\alpha\right]^{-1}$$

Here, P implies predicted value and O implies observed value for the ith data point. For good predictive models, $d^2$ is close to 1. It is observed that the AR(3) and ARMA(3,1) are producing values far away from 1. However, AR-NN(3) is producing $d^2$ above 0.8. This indicates that the AR-NN(3) is significantly better than the other two models in its ability to predict the monthly SN time series. The values of $d^2$ are presented in figure 2. To further visualize the prediction performance, we produce scatterplots in figure 3 for three of the univariate models under consideration. From the scatterplots it is found that majority of the points are clustered around the linear trend line for AR-NN(3). Whereas, for the other two models, the points are deviated far away from the linear trend line. Furthermore, the AR-NN(3) produces highest value of the coefficient of determination ($> 0.6$). Finally, the supremacy of AR-NN(3) over AR(3) and ARMA(3,1) is established. For a visual presentation of the performance of the AR-NN(3) we present a line diagram (figure 04) exhibiting the observed SN for the test cases as well as those predicted by AR-NN(3).

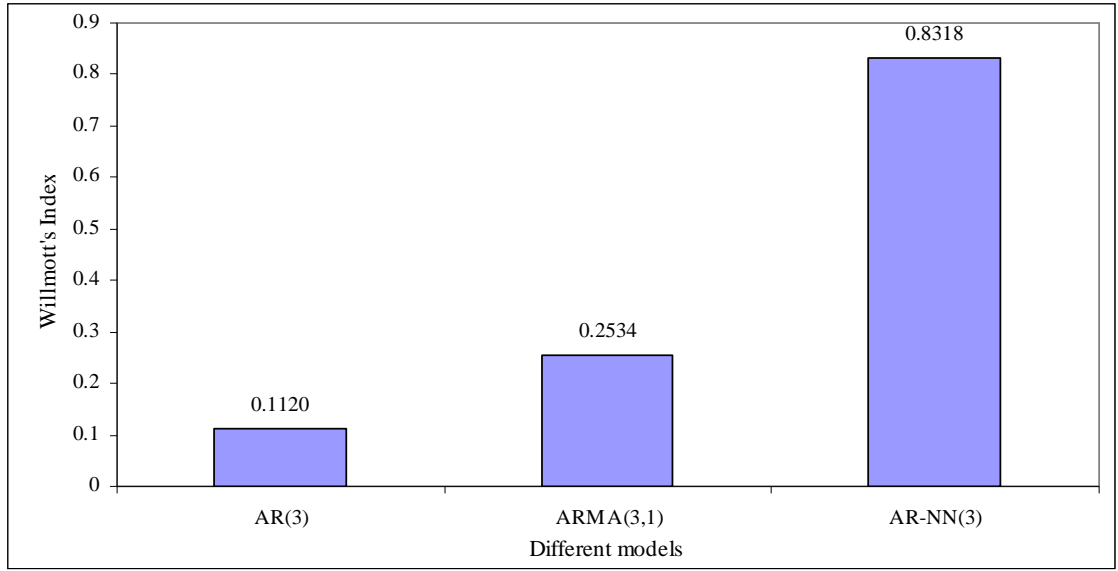

Figure 02. The values of Willmott's index

## 4. Concluding remarks:

In this work we have reported a statistical analysis of monthly sunspot number time series and observed non homogeneity and asymmetry within it. Also, based on Mann-Kendall test a linear trend has been identified. Subsequently, predictive models have been generated in univariate manner. After identifying stationarity within the time series we have generated AR($p$) and ARMA($p,q$). Based on minimization of AIC the best values of $p$ and $q$ have been found to be 3 and 1 respectively. In the next phase, AR-NN(3) has been generated by training a generalized feedforward neural network (GFNN) with two hidden layers and sigmoid nonlinearity. The learning procedure is adopted as Levenberg-Marquardt. Assessing the model performances by means of Willmott's index of second order and coefficient of determination, the performance of AR-NN(3) is identified as better than AR(3) and ARMA(3,1). This study, therefore, establishes the potential of artificial neural network in modeling monthly sunspot number time series and hence identified as a powerful methodology to study the solar activity in statistical manner.

a)

b)

c)

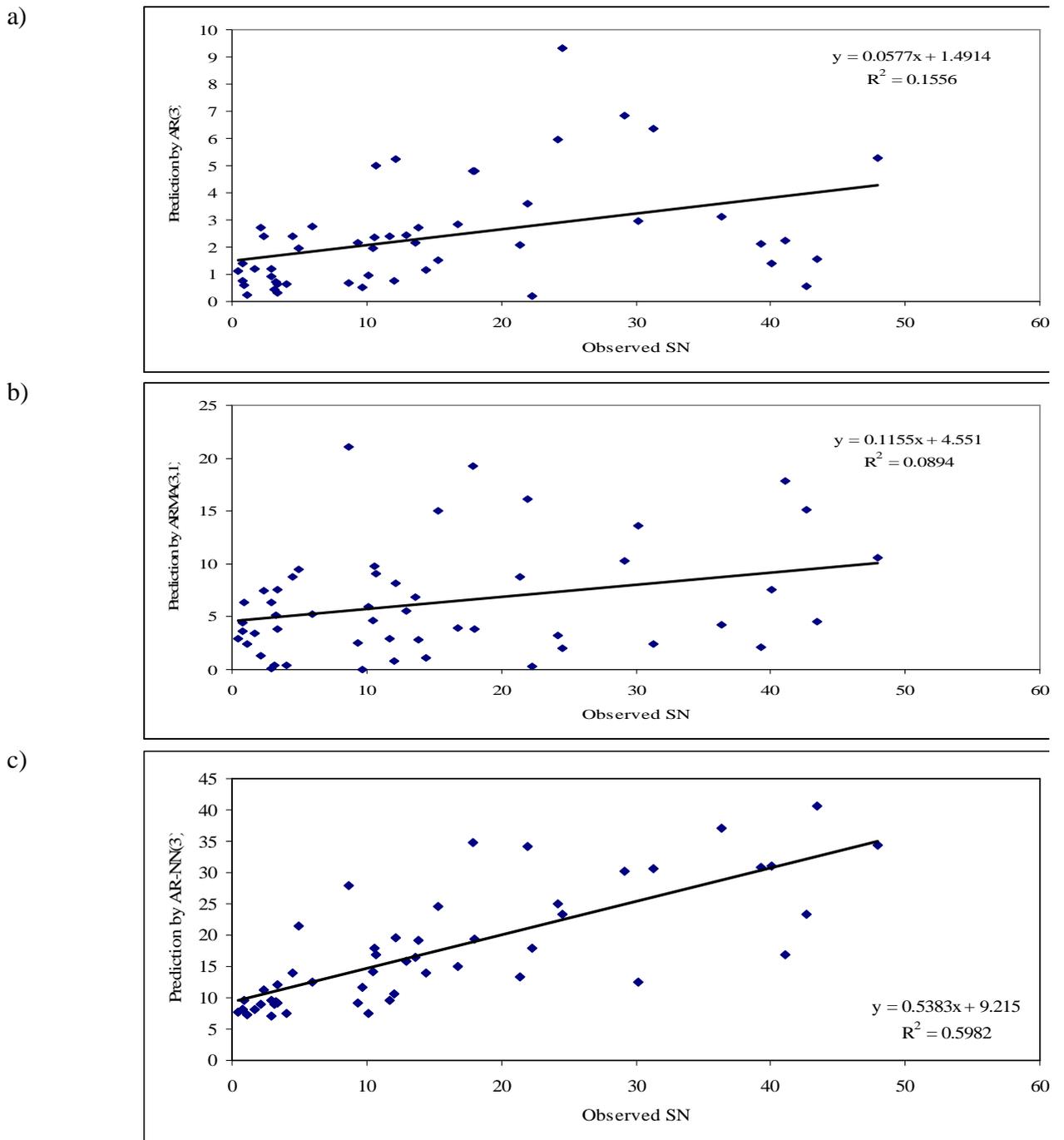

Figure 03 Schematic showing the scatterplots for different models along with the linear trend equation and coefficient of determination $R^2$

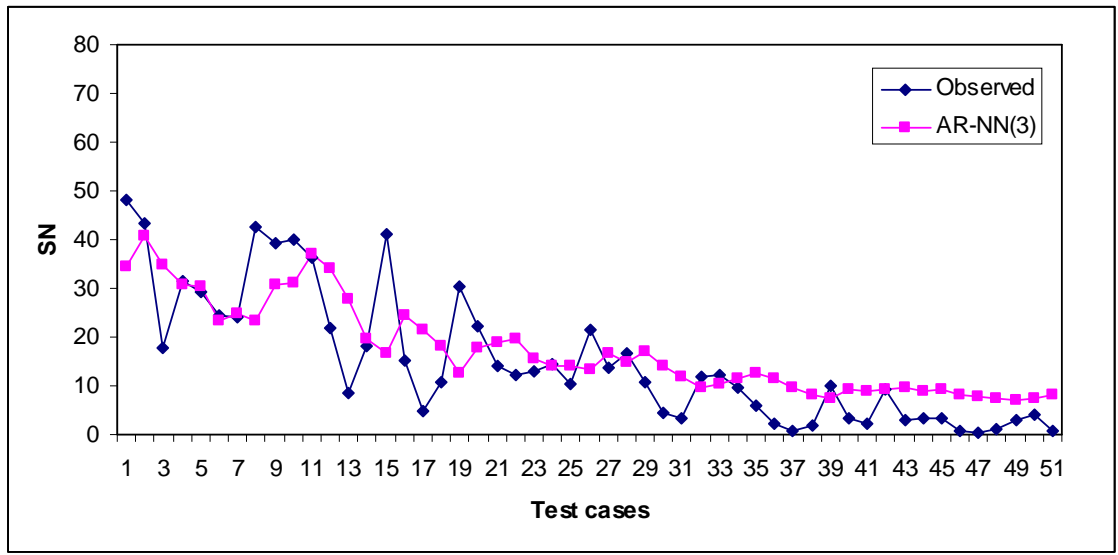

Figure 04. The observed sunspot numbers and those predicted be AR-NN(3).

As future work, we propose to extend the present study to a larger period and to experiment with other neural network methodologies to generate stronger representative models for sunspot numbers.

## Acknowledgements

The authors wish to sincerely acknowledge the warm hospitality provided by Inter-University Centre for Astronomy and Astrophysics (IUCAA), Pune, India, where a major portion of the work was carried out during a scientific visit in January, 2012.